\documentclass[11pt]{article}
\usepackage{amsmath}
\usepackage{amsfonts}
\usepackage{amssymb}
\usepackage{graphicx}

\def\bea{\begin{eqnarray}}
\def\eea{\end{eqnarray}}

\begin{document}
\begin{center}
\LARGE { \bf On the Generalized Minimal Massive Gravity
  }
\end{center}
\begin{center}
{\bf M. R. Setare\footnote{rezakord@ipm.ir} }\\
 {Department of Science, Campus of Bijar, University of  Kurdistan  \\
Bijar, IRAN.}
 \\
 \end{center}
\vskip 3cm

\begin{abstract}
In this paper we study the Generalized Minimal Massive Gravity (GMMG) in asymptotically
$AdS_3$ background. The generalized minimal massive gravity theory is realized
by adding the CS deformation term, the higher derivative deformation term, and an extra term
to pure Einstein gravity with a negative cosmological constant. We study the linearized excitations around the
$AdS_3$ background and find that at special point (tricritical) in parameter space the two massive graviton solutions become massless and they
are replaced by two solutions with logarithmic and logarithmic-squared boundary behavior. So it is natural to proposed that GMMG model could also provide
a holographic description for a $3-$rank Logarithmic Conformal Field Theory (LCFT). We calculate the energy of  the linearized gravitons in $AdS_3$ background, and show that the theory is free of negative-energy bulk modes. Then we obtain the central charges of the CFT dual explicitly and show
GMMG also avoids the aforementioned ``bulk-boundary unitarity clash''. After that we show that General Zwei-Dreibein Gravity (GZDG) model can reduce to GMMG model. Finally by a Hamiltonian analysis we show that the GMMG model has
no Boulware-Deser ghosts and this model propagate only two physical modes.

\end{abstract}

\newpage

\section{Introduction}
It is well known that Einstein gravity suffers from the problem that
the theory is nonrenormalizable in four and higher dimensions.
Adding higher derivative terms such as Ricci and scalar curvature
squared terms makes the theory renormalizable at the cost of the
loss of unitarity \cite{1}. In the other hand pure Einstein-–Hilbert gravity in three dimensions exhibits no propagating physical degrees of freedom \cite{2,3}. But adding the gravitational Chern-Simons term produces a
propagating massive graviton \cite{4}. The resulting theory
is called topologically massive gravity (TMG). Including a negative cosmological constant, yields cosmological topologically massive
 gravity (CTMG). In this case the theory exhibits both gravitons and black holes. Unfortunately there is a problem in this model, with the usual
sign for the gravitational constant, the massive excitations of CTMG carry negative energy. In the absence of a cosmological constant, one can
change the sign of the gravitational constant, but if $\Lambda <0$, this will give a negative mass to the BTZ
black hole, so the existence of a stable ground state is in doubt in this model \cite{5}.\\
 A few years ego \cite{6} a new theory of massive gravity (NMG) in three dimensions
has been proposed. This theory is
 equivalent to the three-dimensional
Fierz-Pauli action for a massive spin-2 field at the linearized level. Moreover NMG in contrast with the TMG \cite{4} is parity
invariant. As a result, the gravitons acquire the same mass for both
helicity states, indicating two massive propagating degrees of freedom. One of common aspects in these two theories
is the existence of AdS vacuum solution. So TMG and NMG provide useful models in which to explore
the AdS/CFT correspondence. The conformal boundary of a three-dimensional
asymptotically anti-de Sitter spacetime is a flat two-dimensional cylinder, and the
asymptotic symmetries are described by a pair of Virasoro algebras \cite{7}. So many study have been done along the route of AdS/CFT correspondence in the TMG and NMG setup \cite{8}. Although, it has been shown the
compliance of the NMG with the holographic c-theorem \cite{9,10}, both TMG and NMG have a bulk-boundary
unitarity conflict. In another term either the bulk or the boundary theory is non-unitary, so there is a clash between the positivity of the two Brown-Henneaux boundary $c$ charges
 and the bulk energies \cite{11}. There is this possibility to extend NMG to higher curvature theories. One of these extension
 of NMG has been done by Sinha \cite{9} where he has added the $R^3$ terms to the action. The other modification is the extension to the Born-Infeld type action \cite{12}. But these extensions of NMG  did not solve
the unitary conflict \cite{9,12,13}. The recently constructed Zwei Dreibein Gravity (ZDG) shows that there is a viable
alternative to NMG \cite{14,15}.\\
It is interesting if one combine TMG and NMG as a generalized massive model in 3-dimension, dubbed Generalized
Massive Gravity theory (GMG), this work first introduced in \cite{6}, then studied more in \cite{16}. This theory has two
mass parameters and TMG and NMG are just two different limits of this generalized
theory.\\
 Recently an interesting three dimensional massive gravity introduced by Bergshoeff, et. al \cite{17} which dubbed Minimal Massive Gravity (MMG), which has the
same minimal local structure as TMG. The MMG model has the same gravitational
degree of freedom as the TMG has and the linearization of the metric field equations for MMG yield a single propagating
massive spin-2 field. It seems that the single massive degree of freedom of MMG is
unitary in the bulk and gives rise to a unitary CFT on the boundary. During last months some interesting works have been done on MMG model \cite{18}.\\
In this paper we would like to unify MMG and NMG into a General Minimal Massive Gravity theory (GMMG).
 The generalized minimal massive gravity theory is realized
by adding the CS deformation term, the higher derivative deformation term, and an extra term
to pure Einstein gravity with a negative cosmological constant. In the other term we would like to extend Generalized
Massive Gravity theory (GMG), by adding an extra term. This theory is expected to have more interesting physics because we can have
one more adjustable mass parameter.\\
Our paper is organized as follows. In section 2 we introduce the GMMG in
$AdS_3$ space. In section 3 we study the linear perturbation around $AdS_3$ vacuum. Then we obtain the solutions of linearized equation
of motion in terms of representations of isometry group $SL(2,R)_{L}\times SL(2,R)_{R}$ of $AdS_{3}$ space. In section 4 we calculate the energy of  the linearized gravitons in $AdS_3$ background. We obtain a condition which is necessary in order the bulk graviton modes have positive energy.
 Then in section 5,  at first we calculate the central charge of CFT dual of the model explicitly.  After that we show that GMMG exhibits
not only massless graviton solutions, but also $log$ and $log^2$ solutions. The $log$ and $log^2$ modes appear in tricritical points in the parameter
space of GMMG model. So in such special point in parameter space all massive gravitons become massless. In another terms the massive graviton modes
that satisfy Brown-Henneaux boundary conditions, in the tricritical points replaced by $log$ and $log^2$ solutions, which obey $log$ and $log^2$
boundary conditions toward $AdS_3$ boundary exactly as what
 occur in GMG model \cite{16}. Therefore it is natural to proposed that GMMG model could also provide
a holographic description for a $3-$rank Logarithmic Conformal Field Theory (LCFT). In section 6 we study the relation between our model and General Zwei-Dreibein Gravity (GZDG) model. In another term we show that GMMG model  can be obtained from Zwei-Dreibein Gravity (ZDG) plus a Lorentz Chern-Simons term for one of the
two spin-connections. Then in section 7 we study the Hamiltonian analysis of the GMMG model and show that there is not
any Boulware-Deser ghosts in the framework of this model. We conclude in section 8 with a discussion of the our results.

\section{The Generalized Minimal Massive Gravity}
The Lagrangian 3-form of MMG is given by \cite{17}
\begin{equation} \label{1}
 L_{MMG}=L_{TMG}+\frac{\alpha}{2}e.h\times h
\end{equation}
where $L_{TMG}$ is the Lagrangian of TMG,
 \begin{equation} \label{1'}
 L_{TMG}=-\sigma e.R+\frac{\Lambda_0}{6} e.e\times e+h.T(\omega)+\frac{1}{2\mu}(\omega.d\omega+\frac{1}{3}\omega.\omega\times\omega)
\end{equation}
 where $\Lambda_0$ is a cosmological parameter with dimension of mass squared, and $\sigma$ a sign. $\mu$ is mass parameter of Lorentz Chern-Simons term. $\alpha$ is a dimensionless parameter, $e$ is dreibein, $h$ is the auxiliary field, $\omega$ is dualised  spin-connection, $T(\omega)$ and
  $R(\omega)$ are Lorentz covariant torsion and curvature 2-form respectively. Now we introduce the Lagrangian
of GMMG model as
  \begin{equation} \label{2}
 L_{GMMG}=L_{GMG}+\frac{\alpha}{2}e.h\times h
\end{equation}
where
\begin{equation} \label{2'}
 L_{GMG}=L_{TMG}-\frac{1}{m^2}(f.R+\frac{1}{2}e.f\times f)
\end{equation}
here $m$ is mass parameter of NMG term and $f$ is an auxiliary one-form field. One can rewrite the Lagrangian 3-form $L_{GMMG}$ as following
\begin{eqnarray*} \label{22'}
 L_{GMMG}&=&-\sigma e_a R^a+\frac{\Lambda_0}{6}\varepsilon^{abc}e_a e_b e_c+h_a T^a+\frac{1}{2\mu}[\omega_a d\omega^a+\frac{1}{3}\varepsilon^{abc}
 \omega_a \omega_b \omega_c]\\
 &-&\frac{1}{m^2}[f_aR^a+\frac{1}{2}\varepsilon^{abc}e_a f_b f_c]+\frac{\alpha}{2}\varepsilon^{abc}e_a h_b h_c
\end{eqnarray*}
The equations of motion of the above Lagrangian by making variation with respect to the fields $h$, $e$, $\omega$ and $f$ are as following respectively

\begin{equation} \label{EOM1}
T(\omega)+\alpha e\times h=0
\end{equation}
\begin{equation} \label{EOM2}
-\sigma R(\omega)+\frac{\Lambda_0}{2}e\times e+D(\omega)h-\frac{1}{2m^2}f\times f+\frac{\alpha}{2}h\times h=0
\end{equation}
\begin{equation} \label{EOM3}
R(\omega)+\mu e\times h-\sigma\mu T(\omega)-\frac{\mu}{m^2}(df+\omega\times f)=0
\end{equation}
\begin{equation} \label{EOM4}
R(\omega)+e\times f=0
\end{equation}
where the locally Lorentz covariant torsion and curvature 2-forms are
\begin{equation} \label{EOM5}
T(\omega)=de+\omega\times e, \hspace{0.5 cm} R(\omega)=d\omega+\frac{1}{2}\omega\times\omega
\end{equation}
The covariant exterior derivative $D(\omega)$ in Eq.(\ref{EOM2}) is given by
\begin{equation} \label{EOM6}
D(\omega)h=dh+\omega\times h
\end{equation}

So by adding extra term $\frac{\alpha}{2}e.h\times h$ to the Lagrangian of generalized massive gravity we obtain Lagrangian of our model.
The equation for metric can be obtained by generalizing  field equation of MMG. Due to this we introduce GMMG field equation
as follows \footnote{Very recently Tekin has introduced another extension of TMG \cite{tek1} like MMG, but it has two massive mode instead of a single one. He has done this extension by introducing following tensor
\begin{equation}\label{f1}
H^{mn}=\frac{1}{2}\eta^{mpq}\nabla_{p}C^{n}_{q}+\frac{1}{2}\eta^{npq}\nabla_{p}C^{m}_{q}
\end{equation}
where $\nabla_{m}H^{mn}=-\nabla_{m}J^{mn}$. So we obtain following Bianchi identity valid for all smooth metrics,
 \begin{equation}\label{f2}
\nabla_{m}(J^{mn}+H^{mn})=0.
\end{equation}
 Due to this, one can obtain $K^{mn}=J^{mn}+H^{mn}$ from the variation of an action. As Tekin \cite{tek1} has mentioned, this action is the quadratic part of the NMG action, which in the first order formalism is given by second term of Eq.(\ref{2'}). Therefore by adding to the field equation of TMG, the tensor  $K^{mn}=J^{mn}+H^{mn}$, one can obtain the NMG deformation of TMG, which is GMG. Our deformation of TMG in this paper is given by adding to the field equation of TMG, the tensor $a_1J^{mn}+a_2H^{mn}$, where $a_1,a_2$ are constant coefficients.}

\begin{equation}
\bar{\sigma}G_{mn}+\bar{\Lambda}_{0}g_{mn}+\frac{1}{\mu }C_{mn}+\frac{\gamma
}{\mu ^{2}}J_{mn}+\frac{s}{2m^{2}}K_{mn}=0,  \label{EOM}
\end{equation}
where
\begin{eqnarray*}
G_{mn} &=&R_{mn}-\frac{1}{2}Rg_{mn},~\ C_{mn}=\epsilon _{m}^{~\ ab}\nabla
_{a}(R_{bn}-\frac{1}{4}g_{bn}R) \\
J_{mn} &=&R_{m}^{\ \ a}R_{an}-\frac{3}{4}RR_{mn}-\frac{1}{2}%
g_{mn}(R^{ab}R_{ab}-\frac{5}{8}R^{2}), \\
K_{mn} &=&-\frac{1}{2}\nabla ^{2}R~g_{mn}-\frac{1}{2}\nabla _{m}\nabla
_{n}R+2\nabla ^{2}R_{mn}+4R_{manb}R^{ab} \\
&-&\frac{3}{2}RR_{mn}-R_{ab}R^{ab}g_{mn}+\frac{3}{8}R^{2}g_{mn},
\end{eqnarray*}%
where $s$ is sign, $\gamma$, $\bar{\sigma}$ $\bar{\Lambda}_{0}$ are the parameters
 which defined in terms of cosmological constant $\Lambda=\frac{-1}{l^2}$, $m$, $\mu$, and the sign of Einstein-Hilbert term.
Here $G_{mn}$ and $C_{mn}~$denote Einstein tensor and Cotton tensor
respectively. Symmetric tensors $J_{mn}~$and $K_{mn}$ are coming from MMG
and NMG parts respectively \cite{6,17}.

\subsection{\protect\bigskip AdS$_{3}~$Solution}

The field equation (\ref{EOM}) admits $AdS_{3}$ solution,
\[
d\bar{s}=\bar{g}_{mn}dx^{m}dx^{n}=l^{2}(-\cosh ^{2}\rho ~d\tau ^{2}+\sinh
^{2}\rho ~d\phi ^{2}+d\rho ^{2})
\]
where $l^{2}\equiv -\Lambda ^{-1}$ fixes with parameters of theory. To show
this consider the Ricci tensor, Ricci scalar\ and Einstein tensor of $%
AdS_{3} $ are
\[
\bar{R}_{mn}=2\Lambda \bar{g}_{mn},~~\bar{R}=6\Lambda ,~\bar{G}%
_{mn}=-\Lambda \bar{g}_{mn}.
\]
Using these results it is easy to see that
\[
\bar{C}_{mn}=0,~\bar{J}_{mn}=\frac{\Lambda ^{2}}{4}\bar{g}_{mn},~~\bar{K}%
_{mn}=-\frac{\Lambda ^{2}}{2}\bar{g}_{mn}.
\]
Then field equation for $AdS_{3}$ reduces to an quadratic equation for $%
\Lambda $
\begin{equation}
\left( \frac{\gamma }{4\mu ^{2}}-\frac{s}{4m^{2}}\right) \Lambda
^{2}-\Lambda \bar{\sigma}+\bar{\Lambda}_{0}=0.  \label{AdS Constraint}
\end{equation}
So,
\begin{equation}  \label{5}
\Lambda=\frac{(\bar{\sigma}\pm\sqrt{\bar{\sigma}^{2}-\bar{\Lambda}_{0}(\frac{\gamma}{\mu^2}-\frac{s}{m^2})})}{\frac{1}{2}
(\frac{\gamma}{\mu^2}-\frac{s}{m^2})}
\end{equation}

\section{Linearized Field Equation}

In this section we study the linear perturbation around the $AdS_{3}$
spacetime correspondence to propagation of graviton. We take vacuum
background $AdS_{3}$ metric as $\bar{g}_{mn}$ and perturb it with a small
perturbation $h_{mn}$ as
\[
g_{mn}=\bar{g}_{mn}+h_{mn}.
\]%
At the first order the field equation (\ref{EOM}) reduces to
\begin{equation}
\bar{\sigma}G_{mn}^{(1)}+\bar{\Lambda}_{0}h_{mn}+\frac{1}{\mu }C_{mn}^{(1)}+%
\frac{\gamma }{\mu ^{2}}J_{mn}^{(1)}+\frac{s}{2m^{2}}K_{mn}^{(1)}=0,
\label{LinEOM1}
\end{equation}%
where
\begin{eqnarray*}
R_{mn}^{(1)} &=&\frac{1}{2}\left( -\bar{\bigtriangledown}^{2}h_{mn}+\bar{%
\bigtriangledown}^{a}\bar{\bigtriangledown}_{m}h_{an}+\bar{\bigtriangledown}%
^{a}\bar{\bigtriangledown}_{n}h_{am}-\bar{\bigtriangledown}_{m}\bar{%
\bigtriangledown}_{n}h\right) , \\
R^{(1)} &=&\left( g^{mn}R_{mn}\right) ^{(1)}=-\bar{\bigtriangledown}^{2}h+%
\bar{\bigtriangledown}_{m}\bar{\bigtriangledown}_{n}h^{mn}-2\Lambda h, \\
G_{mn}^{(1)} &=&R_{mn}^{(1)}-\frac{1}{2}(Rg_{mn})^{(1)}=R_{mn}^{(1)}-\frac{1%
}{2}\bar{g}_{mn}R^{(1)}-3\Lambda h_{mn}, \\
K_{mn}^{(1)} &=&-\frac{1}{2}\bar{\nabla ^{2}}R^{(1)}\bar{g}_{mn}-\frac{1}{2}%
\bar{\nabla}_{m}\bar{\nabla}_{n}R^{(1)}+2\bar{\nabla ^{2}}%
R_{mn}^{(1)}-4\Lambda \bar{\nabla}^{2}h_{mn}-5\Lambda R_{mn}^{(1)}\\
&+&\frac{3}{2%
}\Lambda R^{(1)}\bar{g}_{mn}
+\frac{19}{2}\Lambda ^{2}h_{mn}, \\
C_{mn}^{(1)} &=&\varepsilon _{m}^{~ab}\bar{\nabla}_{a}(R_{bn}^{(1)}-\frac{1}{%
4}\bar{g}_{bn}R^{(1)}-2\Lambda h_{bn}), \\
J_{mn}^{(1)} &=&-\frac{\Lambda }{2}(R_{mn}^{(1)}-\frac{1}{2}\bar{g}%
_{mn}R^{(1)})-\frac{5}{4}\Lambda ^{2}h_{mn}.
\end{eqnarray*}%
Using the fact that $J_{mn}^{(1)}=-\frac{\Lambda }{2}G_{mn}^{(1)}-\frac{%
\Lambda ^{2}}{4}h_{mn}$, we obtain following linearized field equation
\begin{equation}
\left( \bar{\sigma}-\frac{\gamma \Lambda }{2\mu ^{2}}\right) G_{mn}^{(1)}+(%
\bar{\Lambda}_{0}-\frac{\gamma \Lambda ^{2}}{4\mu ^{2}})h_{mn}+\frac{1}{\mu }%
C_{mn}^{(1)}+\frac{s}{2m^{2}}K_{mn}^{(1)}=0.  \label{LinEOM2}
\end{equation}%
Imposing  Eq.(\ref{AdS Constraint}) this equation reduces to

\begin{equation}\label{175}
\frac{\tilde{\mu}}{\mu }G_{mn}^{(1)}+(\frac{s}{2m^{2}}\frac{\Lambda ^{2}}{2}%
+\Lambda \frac{\tilde{\mu}}{\mu })h_{mn}+\frac{1}{\mu }C_{mn}^{(1)}+\frac{s}{%
2m^{2}}K_{mn}^{(1)}=0,~\ \tilde{\mu}=\bar{\sigma}\mu -\frac{\gamma \Lambda }{%
2\mu }
\end{equation}
or in the following form
\[
G_{mn}^{(1)}+(1+\frac{s}{2\tilde{m}^{2}}\frac{\Lambda }{2})\Lambda h_{mn}+%
\frac{1}{\tilde{\mu}}C_{mn}^{(1)}+\frac{s}{2\tilde{m}^{2}}K_{mn}^{(1)}=0,~~%
\tilde{m}^{2}=\frac{\tilde{\mu}}{\mu }m^{2}
\]%
which is exactly same as linearized field equation of GMG \cite{16} ($s=-1$%
). As mention in \cite{17,18} this shows the MMG locally has the
same degrees of freedom. After fixing gauge as $\bar{\nabla}_{a}h_{n}^{a}=0=h$%
, the linearized field equation of GMMG on $AdS_{3}$ becomes \footnote{Having products of d'Alembertian operators (or more precisely, field equations with more than second order time derivatives) is usually a sign of ghosts. This would be what is sometimes called an Ostrogradski instability. }
\begin{equation}\label{6}
\left( \bar{\nabla}^{2}-2\Lambda \right) \left( \bar{\nabla}^{2}h_{mn}+s%
\frac{\tilde{m}^{2}}{\tilde{\mu}}\varepsilon _{m}^{~~ab~}\bar{\nabla}%
_{a}h_{bn}-(-s\tilde{m}^{2}+\frac{5}{2}\Lambda )h_{mn}\right) =0.
\end{equation}
We can define the following operators which commute with each other as
\begin{equation}
(D^{L/R})_{m}^{n}=\delta_{m}^{n}\pm l\varepsilon_{m}^{an}\bar{\nabla}_{a},
\end{equation}
\begin{equation}
(D^{m_{i}})_{m}^{n}=\delta_{m}^{n}+ \frac{1}{m_{i}}\varepsilon_{m}^{an}\bar{\nabla}_{a}, \hspace{0.3cm} i=1,2.
\end{equation}
The equation of motion (\ref{6}) can then be written as
\begin{equation}\label{9}
(D^{L}D^{R}D^{m_1}D^{m_2}h)_{m}^{n}=0,
\end{equation}
The mass parameters $m_1,m_2$ appearing in (\ref{9}) given by
\begin{equation}\label{10}
m_1=\frac{-s\tilde{m}^{2}}{2\tilde{\mu}}+\sqrt{\frac{1}{2l^2}+\bar{\sigma}s\tilde{m}^{2}+\frac{\tilde{m}^{4}}{4\tilde{\mu}^2}}\hspace{0.5cm}
m_2=\frac{-s\tilde{m}^{2}}{2\tilde{\mu}}-\sqrt{\frac{1}{2l^2}+\bar{\sigma}s\tilde{m}^{2}+\frac{\tilde{m}^{4}}{4\tilde{\mu}^2}}
\end{equation}
The GMMG has various critical points in its parameter space where some of differential operator in (\ref{9})degenerate.\\
 Due to the similarity between linearized equation of GMMG and GMG, we can use the result of \cite{16} to find the solution of (\ref{9})
  in terms of representations of isometry group $SL(2,R)_{L}\times SL(2,R)_{R}$ of $AdS_{3}$ space. So one can write the Laplacian acting on tensor $h_{mn}$ in terms of the sum of Casimir operators of $SL(2,R)_{L}$ and $SL(2,R)_{R}$, \cite{19}(see also \cite{20,21}).
\begin{equation}\bar{\nabla}^{2}h_{mn}=-[\frac{2}{l^{2}}(L^{2}+\bar{L}^{2})+\frac{6}{l^{2}}]h_{mn}\end{equation}
Consider states with weight $(h, \bar{h}$:
\begin{equation}
L_0|\psi_{mn}\rangle=h|\psi_{mn}\rangle, \hspace{0.5cm}\bar{L}_0|\psi_{mn}\rangle=\bar{h}|\psi_{mn}\rangle
\end{equation}
Since $|\psi_{mn}\rangle$ are primary states:
\begin{equation}
L_1|\psi_{mn}\rangle=\bar{L}_1|\psi_{mn}\rangle=0.
\end{equation}
For highest weight states, $L^{2}|\psi_{mn}\rangle=-h(h-1)|\psi_{mn}\rangle$. Then for the massless modes we have \footnote{The massless graviton in three dimensions has no degrees of freedom, which is why people call three dimensional gravity a topological theory (one can equivalently write it as a Chern-Simons gauge theory). However, imposing suitable boundary conditions can lead to asymptotically defined global charges which can differ from one solution to another. So relevant `physical states' in 3D GR are characterized by their boundary charge and that's why people sometimes call them boundary gravitons.
 Also in the higher-derivative theories one gets in addition to the massless graviton (which is pure gauge), several massive gravitons which all have 2 degrees of freedom each.}
\begin{equation}h(h-1)+\bar{h}(\bar{h}-1)-2=0,\end{equation}for $h=2+\bar{h}$, we obtain \begin{equation}\bar{h}=\frac{-1\pm1}{2},\,\,\,\,\,\,\,\,\,   h=\frac{3\pm1}{2}\end{equation}but for $h=-2+\bar{h}$ , we have \begin{equation}\bar{h}=\frac{3\pm1}{2},\,\,\,\,\,\,\,\,\,    h=\frac{-1\pm1}{2}.\end{equation}
Also for massive mode simply for $h=2+\bar{h}$ we obtain
\begin{equation}
\bar{h}=\frac{-2+\frac{s\tilde{m}^2l}{\tilde{\mu}}\pm\sqrt{2-4s\tilde{m}^2l^2+\frac{s^2\tilde{m}^4l^2}{\tilde{\mu}^{2}}}}{4}\hspace{0.3cm}
h=\frac{6+\frac{s\tilde{m}^2l}{\tilde{\mu}}\pm\sqrt{2-4s\tilde{m}^2l^2+\frac{s^2\tilde{m}^4l^2}{\tilde{\mu}^{2}}}}{4}
\end{equation}
and for $h=-2+\bar{h}$ , we have
\begin{equation}
\bar{h}=\frac{6-\frac{s\tilde{m}^2l}{\tilde{\mu}}\pm\sqrt{2-4s\tilde{m}^2l^2+\frac{s^2\tilde{m}^4l^2}{\tilde{\mu}^{2}}}}{4}\hspace{0.3cm}
h=\frac{-2-\frac{s\tilde{m}^2l}{\tilde{\mu}}\pm\sqrt{2-4s\tilde{m}^2l^2+\frac{s^2\tilde{m}^4l^2}{\tilde{\mu}^{2}}}}{4}
\end{equation}
The solutions with the
  lower sign will blow up at infinity, thus we consider only the ones with the upper sign.
\section{The energy of linearized gravitons  }
 Now we would like to calculate the energy of the linearized gravitons in $AdS_{3}$ background. The fluctuation $h_{\mu\nu}$ can be decomposed as\cite{19,40}
 \begin{equation}\label{100}
 h_{\mu\nu}=h_{\mu\nu}^{m_1}+h_{\mu\nu}^{m_2}+h_{\mu\nu}^{L}+h_{\mu\nu}^{R}
 \end{equation}
 Using the first order of equation of motion the quadratic action of $h_{\mu\nu}$ is given by

 $$I_{2}=\frac{1}{8\pi G}\int d^{3}x\sqrt{-g}[-(\bar{\sigma}-\frac{\gamma \Lambda}{2\mu^{2}}-\frac{5s\Lambda}{2m^2})\bar{\nabla}_{\lambda}h_{\mu\nu}\bar{\nabla}^{\lambda}h^{\mu\nu}-2\Lambda(\bar{\sigma}-\frac{\gamma \Lambda}{2\mu^{2}}-\frac{5s\Lambda}{2m^2})h_{\mu\nu}h^{\mu\nu}$$
 \begin{equation}\label{101}
 -\frac{1}{\mu}\varepsilon^{\mu\alpha}_{\beta}(\bar{\Box}h^{\beta\nu}-2\Lambda
 h^{\beta\nu})\bar{\nabla}_{\alpha}h_{\mu\nu}+\frac{s}{m^2}\bar{\Box}h_{\mu\nu}(\bar{\Box}h^{\mu\nu}-2\Lambda h^{\mu\nu})]
\end{equation}
  The momentum conjugate to $h_{\mu\nu}$ is
  \begin{equation}\label{102}
  \Pi^{(1)\mu\nu}=\frac{\delta \pounds}{\delta(\bar{\nabla}_{0}h_{\mu\nu})}=\frac{\sqrt{-g}}{8\pi G}[\bar{\nabla}^{0}(-2(\bar{\sigma}-\frac{\gamma \Lambda}{2\mu^{2}}-\frac{3s\Lambda}{2m^2})h^{\mu\nu}+\frac{1}{\mu}\varepsilon_{\beta}^{\alpha\mu}\bar{\nabla}_{\alpha}h^{\beta\nu})]
  \end{equation}
  here $\pounds$ is Lagrangian density.
  Now using the equation of motion we can obtain following expression for momentum conjugate to the left and right modes $h^{\mu\nu}_{L/R}$ and massive modes $h^{\mu\nu}_{m_i}$
  respectively
  \begin{equation}\label{103}
  \Pi_{L/R}^{(1)\mu\nu}=\frac{-\sqrt{-g}}{4\pi G}[\bar{\sigma}+\frac{\gamma }{2\mu^{2}l^2}+\frac{3s}{2m^2l^2}\mp \frac{1}{2\mu l}]\bar{\nabla}^{0}h^{\mu\nu}_{L/R},
  \end{equation}
  \begin{equation}\label{104}
  \Pi_{m_i}^{(1)\mu\nu}=\frac{-\sqrt{-g}}{4\pi G}[\bar{\sigma}+\frac{\gamma }{2\mu^{2}l^2}+\frac{3s}{2m^2l^2}- \frac{m_i}{2\mu l}]\bar{\nabla}^{0}h^{\mu\nu}_{m_i}
  \end{equation}
  since we have up to four time derivatives in the lagrangian, one can introduce a canonical variable using the Ostrogradsky method \cite{19,40} as $K_{\mu\nu}=\bar{\nabla}_{0}h_{\mu\nu}$. The conjugate momentum of this variable is given
  \begin{equation}\label{105}
  \Pi^{(2)\mu\nu}=\frac{\sqrt{-g}g^{00}}{8\pi G}[\frac{-1}{\mu}\varepsilon^{\rho\alpha\mu}\bar{\nabla}_{\alpha}h^{\nu}_{\rho}+\frac{2s\Lambda}{m^2}h^{\mu\nu}]
  \end{equation}
then using the equations of motion we obtain
\begin{equation}\label{106}
\Pi_{L/R}^{(2)\mu\nu}=\frac{-\sqrt{-g}g^{00}}{4\pi G}(\frac{s}{m^2l^2}\mp\frac{1}{2\mu l})h^{\mu\nu}_{L/R},
\end{equation}
\begin{equation}\label{107}
\Pi_{m_i}^{(2)\mu\nu}=\frac{-\sqrt{-g}g^{00}}{4\pi G}(\frac{s}{m^2l^2}-\frac{m_i}{2\mu })h^{\mu\nu}_{m_i},
\end{equation}
Now we can write the Hamiltonian of the system as
  \begin{equation}
  H=\int d^{2}x(\dot{h}_{\mu\nu}\Pi^{(1)\mu\nu}+\dot{K}_{\mu\nu}\Pi^{(2)\mu\nu}-\pounds),
  \end{equation}\label{108}
Then by substituting the equation of motion of the highest weight states, we obtain the energy of left and right modes $h^{\mu\nu}_{L/R}$ and massive modes $h^{\mu\nu}_{m_i}$ as following

$$E_{L/R}=\int d^{2}x\,(\dot{h}_{L/R,\mu\nu}\Pi^{(1)\mu\nu}_{L/R}+\dot{K}_{L/R,\mu\nu}\Pi^{(2)\mu\nu}_{L/R}-\pounds)$$
\begin{equation}\label{109}
=\frac{-1}{4\pi G} (\bar{\sigma}+\frac{\gamma }{2\mu^{2}l^2}+\frac{s}{2m^2l^2})\int d^{2}x\ \sqrt{-g}\bar{\nabla}^{0}h_{\mu\nu}^{L/R}\dot{h}^{\mu\nu}_{L/R}
 \end{equation}
$$E_{m_i}=\int d^{2}x\,(\dot{h}_{m_i,\mu\nu}\Pi^{(1)\mu\nu}_{m_i}+\dot{K}_{m_i,\mu\nu}\Pi^{(2)\mu\nu}_{m_i}-\pounds)$$
\begin{equation}\label{110}
=\frac{-\sqrt{-g}}{4\pi G} (\bar{\sigma}+\frac{\gamma }{2\mu^{2}l^2}+\frac{s}{2m^2l^2})\int d^{2}x\ \sqrt{-g}\bar{\nabla}^{0}h_{\mu\nu}^{m_i}\dot{h}^{\mu\nu}_{m_i}
 \end{equation}
 Since the integrals in Eqs.$(\ref{109})$ and $(\ref{110})$ are negative, therefore we find that if $\bar{\sigma}+\frac{\gamma }{2\mu^{2}l^2}+\frac{s}{2m^2l^2}>0$ then the energy of left, right and also massive modes are positive. This is no-ghost condition in the framework of GMMG model, and under this condition the theory is free of negative-energy bulk modes. In the next section we calculate the central charges of dual CFT explicitly and show that the above condition is consistent with the requirement of positive central charges.
\section{Central charges and logarithmic modes }
Now we obtain the  central charges of the CFT dual of GMMG. We consider following Brown-Henneux boundary condition for the linearized gravitational excitation in asymptotically $AdS_3$ spacetime in the global coordinate system, as has been done in \cite{16}.
\[ \left( \begin{array}{ccc}
 f_{++}& f_{+-} &e^{-2\rho} \\
 f_{+-}& f_{--} & e^{-2\rho}  \\
 e^{-2\rho}& e^{-2\rho} &e^{-2\rho}\\
 \end{array} \right)\]
The corresponding asymptotic Killing vectors are given by
\begin{eqnarray*}
\xi &=&[\epsilon^{+}(\tau^{+})+2e^{-2\rho}\partial_{-}^{2}\epsilon^{-}(\tau^{-})+e^{-4\rho}]\partial_{+}\\
&+&[\epsilon^{-}(\tau^{-})+2e^{-2\rho}\partial_{+}^{2}\epsilon^{+}(\tau^{+})+e^{-4\rho}]\partial_{-}\\
&-&\frac{1}{2}[\partial_{+}\epsilon^{+}(\tau^{+})+\partial_{-}\epsilon^{-}(\tau^{-})+e^{-2\rho}]\partial_{\rho}
\end{eqnarray*}%
where $\tau^{\pm}=\tau\pm\phi$, $\epsilon^{+}_{m}=e^{im\tau^{+}}$ and $\epsilon^{-}_{n}=e^{in\tau^{-}}$.\\
Very recently the conserved charges of GMMG in asymptotically $AdS_3$ spacetime have been obtained in \cite{set2}. According to the results of \cite{set2}, conserved charge is given by
\begin{equation}\label{300}
Q^{\mu} (\bar{\xi})=\frac{c}{16 \pi G} \int_{\Sigma}  dl_{i} \left[ \left( \bar{\sigma} -\frac{\gamma \Lambda}{2 \mu ^{2}} - \frac{s \Lambda}{2m^{2}} \right) q_{E}^{\mu i} (\bar{\xi})+\frac{1}{2 \mu} q_{E}^{\mu i} (\bar{\Xi}) + \frac{1}{2 \mu} q_{C}^{\mu i} (\bar{\xi}) +\frac{s}{2m^{2}} q_{N}^{\mu i} (\bar{\xi}) \right],
\end{equation}
where
$$ q_{E}^{\mu \nu} (\bar{\xi}) = 2 \sqrt{-\bar{g}} \left( \bar{\xi}_{\lambda}  \bar{\nabla}^{ [ \mu} h^{\nu ] \lambda}
 + \bar{\xi}^{ [ \mu}  \bar{\nabla}^{\nu ] } h + h^{\lambda [ \mu}  \bar{\nabla}^{\nu ] } \bar{\xi}_{\lambda} +
\bar{\xi} ^{ [ \nu} \bar{\nabla} _{ \lambda} h^{\mu ] \lambda} + \frac{1}{2} h \bar{\nabla} ^{ \mu} \bar{\xi}^{\nu} \right) , $$
$$ q_{C}^{\mu \nu} (\bar{\xi}) = \varepsilon ^{\mu \nu}  _{\hspace{3 mm} \alpha} \mathcal{G}_{L}^{\alpha \beta} \bar{\xi}_{\beta}
 + \varepsilon ^{\beta \nu} _{\hspace{3 mm} \alpha} \mathcal{G}_{L}^{\mu \alpha} \bar{\xi}_{\beta}
 +\varepsilon ^{\mu \beta}  _{\hspace{3 mm} \alpha} \mathcal{G}_{L}^{\alpha \nu} \bar{\xi}_{\beta}, $$
 \begin{equation}\label{280}
q_{N}^{\mu \nu} (\bar{\xi}) = \sqrt{-\bar{g}} \left[ 4 \left(  \bar{\xi}_{\lambda}  \bar{\nabla}^{ [ \nu} \mathcal{G}_{L}^{\mu ] \lambda}
+\mathcal{G}_{L}^{\lambda [ \nu} \bar{\nabla} ^{ \mu ] } \bar{\xi}_{\lambda} \right) + \bar{\xi}^{ [ \mu}  \bar{\nabla}^{\nu ] } R_{L}
+ \frac{1}{2} R_{L} \bar{\nabla} ^{ \mu} \bar{\xi}^{\nu} \right] ,
\end{equation}
also, $ \bar{\Xi}^{\beta}=\frac{1}{\sqrt{-\bar{g}}} \varepsilon^{\alpha \lambda \beta} \bar{\nabla} _{ \alpha} \bar{\xi}^{\lambda}$. By substituting the linearized gravitational excitation into Eq.(\ref{280}) and finally in Eq.(\ref{300}), we obtain following expression for conserved charge in the limit $\rho\rightarrow \infty$

$$Q^{0} (\bar{\xi})=\frac{-c}{16 \pi lG} \int _{\bar{\sigma}}  dl_{i} [ \left( \bar{\sigma} +\frac{\gamma }{2 \mu ^{2}l^2} + \frac{s }{2m^{2}l^2}+\frac{1}{\mu l} \right)\epsilon^{+}f_{++} +\left( \bar{\sigma} +\frac{\gamma }{2 \mu ^{2}l^2} + \frac{s }{2m^{2}l^2}-\frac{1}{\mu l} \right)\epsilon^{-}f_{--} +$$
\begin{equation}\label{180}
\left( \bar{\sigma} +\frac{\gamma }{2 \mu ^{2}l^2} + \frac{s }{2m^{2}l^2} \right)\frac{(\epsilon^{+}+\epsilon^{-})(16f_{+-}-f_{\rho\rho})}{16}]
\end{equation}
By substituting Brown-Henneux boundary condition for the linearized gravitational excitation into linearized
field equation (\ref{LinEOM2}), in the limit $\rho\rightarrow\infty$ one can obtain from the $\rho\rho$ component that
\begin{equation}\label{181}
16f_{+-}-f_{\rho\rho}=0
\end{equation}
Imposing the above equation on the above conserved charge $Q^{0} (\bar{\xi})$, we obtain
 $$Q^{0} (\bar{\xi})=\frac{-c}{16 \pi lG} \int _{\bar{\sigma}}  dl_{i} [ \left( \bar{\sigma} +\frac{\gamma }{2 \mu ^{2}l^2} + \frac{s }{2m^{2}l^2}+\frac{1}{\mu l} \right)\epsilon^{+}f_{++} +\left( \bar{\sigma} +\frac{\gamma }{2 \mu ^{2}l^2} + \frac{s }{2m^{2}l^2}-\frac{1}{\mu l} \right)\epsilon^{-}f_{--}] $$
\begin{equation}\label{182}
=Q_R+Q_L
\end{equation}
where $Q_L$ and $Q_R$ are left moving and right moving conserved charges respectively,
\begin{equation}\label{183}
Q_R=\frac{-c}{16 \pi lG} \int _{\Sigma}  dl_{i} \left( \bar{\sigma} +\frac{\gamma }{2 \mu ^{2}l^2} + \frac{s }{2m^{2}l^2}+\frac{1}{\mu l} \right)\epsilon^{+}f_{++}
\end{equation}

\begin{equation}\label{183}
Q_L=\frac{-c}{16 \pi lG} \int _{\Sigma}  dl_{i} \left( \bar{\sigma} +\frac{\gamma }{2 \mu ^{2}l^2} + \frac{s }{2m^{2}l^2}-\frac{1}{\mu l} \right)\epsilon^{-}f_{--}.
\end{equation}
These conserved charges satisfy two copies of Virasoro algebra with following left and right central charges

\begin{equation}\label{18}
c_{R}=\frac{3l}{2G}\left( \bar{\sigma} +\frac{\gamma }{2 \mu ^{2}l^2} + \frac{s }{2m^{2}l^2}+\frac{1}{\mu l} \right)%
,~~c_{L}=\frac{3l}{2G} \left( \bar{\sigma} +\frac{\gamma }{2 \mu ^{2}l^2} + \frac{s }{2m^{2}l^2}-\frac{1}{\mu l} \right),
\end{equation}
where we have considered arbitrary coefficient $c=-2$.
So the asymptotic symmetry algebra of $AdS_3$ space in GMMG model consists of two copies of the Virasoro algebra with the above central charges. If $\bar{\sigma}=\sigma$, the above cental charges in the limit $\frac{1}{m^2}\rightarrow 0$ reduce to the result for MMG \cite{tek2}, and in the limit $\gamma\rightarrow 0$, $s=-1$, reduce to the cental charges  for GMG \cite{16}. So in order we have consistent model with previous massive gravity models in 3-dimension, the parameter $\bar{\sigma}$ in mentioned limiting cases should reduce to the usual parameter $\sigma$.
In the previous section we have determined the condition that the graviton bulk modes are not ghost, i.e. $\bar{\sigma}+\frac{\gamma }{2\mu^{2}l^2}+\frac{s}{2m^2l^2}>0$. Since $\mu l$  is positive, by mentioned condition we obtain positive value for right cental charge, so $ \bar{\sigma} +\frac{\gamma }{2 \mu ^{2}l^2} + \frac{s }{2m^{2}l^2}+\frac{1}{\mu l} >0$. If $ \bar{\sigma} +\frac{\gamma }{2 \mu ^{2}l^2} + \frac{s }{2m^{2}l^2}>\frac{1}{\mu l} $, then the left central charge is also positive. This condition is not contradict with previous no-ghost condition. So GMMG also avoids the aforementioned ``bulk-boundary unitarity clash''. Due to these we have a semi-classical quantum gravity model in $2+1$ dimension which in both bulk and boundary is unitary, so is a consistent model. \\
At the critical line,
\begin{equation}\label{20}
 \bar{\sigma}=\frac{1}{\mu l}-\frac{s}{2m^{2}l^{2}}-\frac{\gamma }{2 \mu ^{2}l^2}
\end{equation}
 we have
 \begin{equation}\label{21}
 c_{L}=0, \hspace{0.5cm} c_{R}=\frac{3}{G\mu}.
\end{equation}
 In this case the linearized equation of motion (\ref{9})becomes
 \begin{equation}\label{22}
(D^{R}D^{L}D^{L}D^{m_2}h)_{m}^{n}=0,
\end{equation}
 in another term the operators $D^{m_1}$ and $D^{L}$ degenerate here. Due to this, massive graviton with mass $m_1$ degenerate with left massless graviton. Therefore at $c_{L}=0$ a new logarithmic solution appear. Logarithmic solution satisfies
 \begin{equation}\label{23}
(D^{L}D^{L}h^{log})_{m}^{n}=0, \hspace{0.5cm}(D^{L}h^{log})_{m}^{n}\neq0.
\end{equation}
In GMMG there is another critical line where the operators $D^{m_1}$ and $D^{m_2}$ degenerate. This line can be obtained when $m_1=m_2$:
\begin{equation}\label{24}
\bar{\sigma}=\frac{-s m^{2}}{4\mu^2}-\frac{1}{2s\tilde{m}^{2}l^2}
\end{equation}
At the intersection of the critical line (\ref{20}) and (\ref{24}) one can see a critical point as
 \begin{equation}\label{25}
\frac{1}{\mu l}-\frac{s}{2m^{2}l^{2}}-\frac{\gamma }{2 \mu ^{2}l^2}=\frac{-s m^{2}}{4\mu^2}-\frac{1}{2s\tilde{m}^{2}l^2}
\end{equation}
 where the left central charge $c_{L}=0$, and three operators $D^{L}$, $D^{m_1}$ and $D^{m_2}$ degenerate. Due to this, the intersection of the critical line
 (\ref{20}) and (\ref{24}) is a tricritical point. More than this, there is another tricritical point in GMMG. $c_{R}=0$, give us a new critical line as
 \begin{equation}\label{26}
 \bar{\sigma}=-\left( \frac{\gamma }{2 \mu ^{2}l^2} + \frac{s }{2m^{2}l^2}+\frac{1}{\mu l} \right)
\end{equation}
 By intersecting the above line with line (\ref{24}), we obtain following new tricritical point
 \begin{equation}\label{27}
\frac{\gamma }{2 \mu ^{2}l^2} + \frac{s }{2m^{2}l^2}+\frac{1}{\mu l}=\frac{s m^{2}}{4\mu^2}+\frac{1}{2s\tilde{m}^{2}l^2}
\end{equation}
 In this new tricritical point the operators $D^{m_1}$ and $D^{m_2}$ degenerate with $D^{R}$. So the equations of motion in the first and second mentioned tricritical
 points are given respectively by

 \begin{equation}\label{28}
(D^{R}D^{L}D^{L}D^{L}h)_{m}^{n}=0,
\end{equation}
 \begin{equation}\label{29}
(D^{R}D^{R}D^{R}D^{L}h)_{m}^{n}=0,
\end{equation}
 Due to the tricritical points, more than the logarithmic solutions, we have the square-logarithmic mode $h_{mn}^{log^{2}}$
 \begin{equation}\label{30}
(D^{L}D^{L}D^{L}h^{log^{2}})_{m}^{n}=0, \hspace{0.5cm} , (D^{L}D^{L}h^{log^{2}})_{m}^{n}\neq0
\end{equation}
 Similar to the Eqs.(\ref{23}), (\ref{30}), in second tricritical point we have following equations

  \begin{equation}\label{31}
(D^{R}D^{R}h^{log})_{m}^{n}=0, \hspace{0.5cm}(D^{R}h^{log})_{m}^{n}\neq0.
\end{equation}
 \begin{equation}\label{32}
(D^{R}D^{R}D^{R}h^{log^{2}})_{m}^{n}=0, \hspace{0.5cm} , (D^{R}D^{R}h^{log^{2}})_{m}^{n}\neq0
\end{equation}
The modes $h^{log}_{mn}$, $h^{log^{2}}_{mn}$ in contrast with modes $h^{R}_{mn}$, $h^{L}_{mn}$, do not obey the usual Brown-Henneaux boundary conditions,
 they satisfy a $log$ and $log^{2}$ asymptotic behavior at the boundary \cite{16}.
\\ Similar to what one can obtain in GMG \cite{22,23}, it seems reasonable to conjecture that for GMMG at the critical line (\ref{20}), the dual CFT is a LCFT. So, as GMG \cite{22,23} this kind of degeneration allows for the possibility of an LCFT.
\section{Relation of the GMMG model with GZDG}
The authors of \cite{26'} have obtained the Chern-Simons-like formulation of NMG from ZDG model by field and parameter redefinitions . Similarly
in this section we show that GMMG model  can be obtained from Zwei-Dreibein Gravity (ZDG) plus a Lorentz Chern-Simons term for one of the
two spin-connections. In another term we show General Zwei-Dreibein Gravity (GZDG) \cite{15} can reduce to GMMG. The Lagrangian $3$-form of ZDG is \cite{14} (see also \cite{15})
\begin{equation}\label{35}
L_{ZDG}=-M_P[\sigma e_1.R_1+e_2.R_2+\frac{m^2}{6}(\alpha_1 e_1.e_1\times e_1+\alpha_2 e_2.e_2\times e_2)-\frac{m^2}{2}(\beta_1e_1.e_1\times e_2+\beta_2e_1.e_2\times e_2)]
\end{equation}
where $R_{1}^{a}$ and $R_{2}^{a}$ are the dualised Riemann $2$-forms constructed from $\omega_{1}^{a}$ and $\omega_{2}^{a}$ respectively. Also $\alpha_1$ and $\alpha_2$ are two dimensionless cosmological parameters and $\beta_1$ and $\beta_2$ are two dimensionless coupling constants.
The authors of \cite{15} generalized the above Lagrangian to GZDG model by adding a Lorentz Chern-Simons term as, but with $\beta_2=0$
\begin{equation}\label{36}
L_{GZDG}=L_{ZDG}(\beta_2=0)+\frac{M_P}{2\mu}(\omega_1.d\omega_1+\frac{1}{3}\omega_1.\omega_1\times\omega_1)
\end{equation}
Here we consider the above Lagrangian but with $\beta_2\neq0$. Now we consider following field redefinitions
\begin{equation}\label{37}
e_1\rightarrow e+x f, \hspace{1cm} e_2\rightarrow e
\end{equation}

\begin{equation}\label{38}
\omega_1\rightarrow \omega, \hspace{1cm}\omega_2\rightarrow \omega+y h.
\end{equation}
where $x$ and $y$ are arbitrary parameters. By the above field redefinitions, the Lagrangian (\ref{36}) reduce to the following
\begin{eqnarray*}\label{39}
L_{GZDG}&=&M_P[-(1+\sigma)e.R-m^2(\frac{\alpha_1+\alpha_2}{6}-\frac{\beta_1+\beta_2}{2})e.e\times e\\
&-&[x\sigma f.R+\frac{m^2x^2}{2}(\alpha_1-\beta_1)e.f\times f]-\frac{y^2}{2}e.h\times h\\
&+&\frac{1}{2\mu}(\omega.d\omega+\frac{1}{3}\omega.\omega\times\omega)-ye.D(\omega)h\\
&-&\frac{xm^2}{2}(\alpha_1-2\beta_1-\beta_2)e.e\times f-\frac{\alpha_1 m^2x^3}{6}f.f\times f]
\end{eqnarray*}
As one can see all terms of the GMMG model are generated, plus a couple last terms which are extra terms.
Now by considering $\alpha_1=0$, $\beta_2=-2\beta_1$, we remove these extra terms. Then by following parameter redefinitions
\begin{eqnarray*}\label{40}
1+\sigma\rightarrow\sigma, \hspace{1cm} \frac{m^2x^2}{2}\beta_1\rightarrow\frac{-1}{2m^2} \hspace{1cm} m^2(\frac{\alpha_2}{6}+\frac{\beta_1}{2})\rightarrow \frac{\Lambda_0}{6}\\
\frac{-y^2}{2}\rightarrow\frac{\alpha}{2} \hspace{1cm} -x\sigma\rightarrow \frac{-1}{m^2} \hspace{1cm} -y\rightarrow 1,
\end{eqnarray*}
also since
\begin{equation}\label{41}
ye.D(\omega)h=y h.T,
\end{equation}
the Lagrangian (\ref{36}) reduce to the Lagrangian of GMMG model.
\section{Hamiltonian Analysis}
In this section by a Hamiltonian analysis we show that the GMMG model has
no Boulware-Deser ghosts. In section 2 we have written the Lagrangian of GMMG as a Lagrangian 3-form constructed from one form
fields and their exterior derivatives. So the GMMG model takes a Chern-Simons-like form. As has been discussed in \cite{15} the Chern-Simons formulation of gravity models is well-adapted to a Hamiltonian analysis. It is important that by a Hamiltonian analysis one can obtain the number of local degrees of freedom exactly and independent of a linearised approximation. Following the approach of \cite{15} (see also \cite{25'})) we can rewrite  Lagrangian 3-form of GMMG as
\begin{equation}\label{410}
L=\frac{1}{2}g_{rs}a^{r}.da^{s}+\frac{1}{6}f_{rst}a^{r}.(a^{s}\times a^{t})
\end{equation}
where $g_{rs}$ is a symmetric constant metric on the flavour space which is invertible, so it can be used to raise and lower flavour indices, and the coupling constants $f_{rst}$, which is totally symmetric flavour tensor. GMMG model has four flavours of one-forms: the dreibein $a^{ea}=e^a$, the dualised spin-connection $a^{\omega a}=\omega^a$, and and two extra fields $a^{fa}=f^a$, $a^{ha}=h^a$.\\
By following space-time split
 \begin{equation}\label{810}
a^{ra}=a_{0}^{ra}dt+a_{i}^{ra}dx^{i}
\end{equation}
we can write the Lagrangian density as
\begin{equation}\label{811}
\pounds=\frac{-1}{2}\varepsilon^{ij}g_{rs}a^{r}_{i}.a^{s}_{j}+a_{0}^{r}.\phi_{r}
\end{equation}
where $\varepsilon^{ij}=\varepsilon^{0ij}$. The Lagrange multipliers for the primary constraints $\phi_{a}^{r}$ are $a_{0}^{ra}$, which are the time components of the fields, and
\begin{equation}\label{812}
\phi_{a}^{r}=\varepsilon^{ij}(g_{rs}\partial_{i}a^{sa}_{j}+\frac{1}{2}f_{rst}(a^{s}_{i}\times a^{t}_{j}))
\end{equation}
By comparing Lagrangian 3-form of GMMG which is given under equation(\ref{2'}) with Lagrangian (\ref{410}), we obtain following nonzero components of flavour-space metric
$g_{rs}$ and the structure constants $f_{rst}$,
\begin{eqnarray*}\label{400}
g_{\omega e}=-\sigma, \hspace{0.5cm} g_{eh}=1, \hspace{0.5cm}g_{f\omega}=\frac{-1}{m^2},\hspace{0.5cm} g_{\omega\omega}=\frac{1}{\mu} \\
f_{e\omega\omega}=-\sigma \hspace{0.5cm}f_{e h\omega}=1 \hspace{0.5cm}f_{eff}= \frac{-1}{m^2},\hspace{0.5cm}f_{\omega\omega\omega}= \frac{1}{\mu}\\
f_{\omega\omega f}=\frac{-1}{m^2},\hspace{0.5cm}f_{eee}=\Lambda_0 \hspace{0.5cm}f_{ehh}=\alpha.
\end{eqnarray*}
Now we consider following the integrability conditions
\begin{equation}\label{411}
A_{qrsp}a^{ra}a^p.a^q=0
\end{equation}
where $A_{qrsp}=f^{t}_{q[r}f_{s]pt}$. The consistency of the primary constraints is equivalent to satisfying the integrability conditions (\ref{411}) \cite{15}. Using the above integrability conditions we obtain following 3-form relations,
\begin{equation}\label{412}
f^a[\frac{1}{\mu}e.f+(1+\alpha\sigma)h.e]-(1+\alpha\sigma)h^a e.f+\frac{\alpha}{m^2}f^a f.h=0
\end{equation}
\begin{equation}\label{413}
e^a[\frac{1}{\mu}e.f+(1+\alpha\sigma)h.e+\frac{\alpha}{m^2}f.h]-\frac{\alpha}{m^2}h^a e.f=0
\end{equation}
\begin{equation}\label{414}
(1+\alpha\sigma)e^ae.f-\frac{\alpha}{m^2}f^ae.f=0
\end{equation}
as one expected the above equation reduced to the corresponding relations for GMG in the limit $\alpha\rightarrow 0$, where have been obtained in \cite{15}.\\
The secondary constraints are given by
\begin{equation}\label{415}
\psi_{s}=B_{rs}=f^{t}_{q[r}f_{s]pt}\Delta^{pq}
\end{equation}
where $\Delta^{pq}=\varepsilon^{ij}a_{i}^{p}.a_{j}^{q}$. Now we assume $(1+\alpha\sigma) e^a - \frac{\alpha}{m^2} f^a$ to have an inverse. \footnote{This assumption of invertibility is similar to the the assumed invertibility of $\beta_1e_1 + \beta_2 e_2$ in the (G)ZDG \cite{15}.} By this assumption we restrict our model, such that Boulwar-Deser ghost does not appear as the degrees of freedom of the model. Then from Eq.(\ref{414}) we obtain the following secondary constraint
\begin{equation}\label{416}
e.f=\Delta^{ef}=0,
\end{equation}
then by an appropriate linear combination of Eqs.(\ref{412}) and (\ref{413}) we have
\begin{equation}\label{4160}
e.f[C\frac{f^a}{\mu}-C(1+\alpha\sigma)h^a+D(\frac{e^a}{\mu}-\frac{\alpha}{m^2}h^a) ]+(Cf^a+De^a)[(1+\alpha\sigma)h.e+\frac{\alpha}{m^2}f.h]=0
\end{equation}
where $C$ and $D$ are arbitrary constants, here we assume $C=-\frac{\alpha}{m^2}$ and $D=1+\alpha\sigma$. So by considering secondary constraint (\ref{416}), and since $(1+\alpha\sigma) e^a - \frac{\alpha}{m^2} f^a$ is invertible we derive the second secondary constraint
\begin{equation}\label{4161}
h.[(1+\alpha\sigma)e-\frac{\alpha}{m^2}f]=0
\end{equation}
where again in the limiting case $\alpha\rightarrow 0$, we obtain the two secondary constraint of GMG in \cite{15}.
Now we should obtain the rank of matrix $P_{rs}^{pq}$ which id defined as following
\begin{equation}\label{417}
P_{rs}^{pq}=B_{rs}\eta^{ab}+C_{rs}^{pq}
\end{equation}
where $C_{rs}^{pq}=2A_{rsqp}(V^{ab})^{pq}$, and $V_{ab}^{pq}=\varepsilon^{ij}a_{ia}^{p}.a_{jb}^{q}$. Taking into account the two secondary constraints , the first term in $P_{rs}^{pq}$ omit. So in the basis $(e,\omega,f,h)$, matrix $P_{rs}^{pq}$ takes following form:
\[ \left( \begin{array}{cccc}
\frac{V_{ab}^{ff}}{\mu}-2(1+\alpha\sigma)V_{[ab]}^{fh} & 0 &\frac{-V_{ab}^{fe}}{\mu}+\frac{\alpha}{m^2}V_{ab}^{fh}+(1+\alpha\sigma)V_{ab}^{he}  & (1+\alpha\sigma)V_{ab}^{fe}-\frac{\alpha}{m^2}V_{ab}^{ff} \\
0 & 0 & 0 & 0 \\
\frac{-V_{ab}^{ef}}{\mu}+\frac{\alpha}{m^2}V_{ab}^{hf}+(1+\alpha\sigma)V_{ab}^{eh} & 0 &\frac{V_{ab}^{ee}}{\mu}-\frac{\alpha}{m^2}V_{[ab]}^{eh}  &- (1+\alpha\sigma)V_{ab}^{ee}+\frac{\alpha}{m^2}V_{ab}^{ef} \\
(1+\alpha\sigma)V_{ab}^{ef}-\frac{\alpha}{m^2}V_{ab}^{ff} & 0 & -(1+\alpha\sigma)V_{ab}^{ee}+\frac{\alpha}{m^2}V_{ab}^{fe} & 0 \end{array} \right)\]
when we consider the limit $\alpha\rightarrow 0$, the above matrix reduce to the corresponding matrix for GMG model \cite{15}. \footnote{ Please note that the basis for matrix $P_{rs}^{pq}$ in GMG case in \cite{15} is as $(\omega,h,e,f)$.} The rank of the above matrix at an arbitrary point in space-time is 4. From following equation, one can obtain the dimension of the physical phase space per space point
\begin{equation}\label{418}
D=6N-2(3N-rank P-M)-(rank P+2M)=rank P,
\end{equation}
where $N$ is the number of flavours, and $M$ is the number of secondary constraints. In our case, $N=4$, $rank P=4$, $M=2$. So,
\begin{equation}\label{419}
D=6\times4-2(12-4-2)-(4+4)=4.
\end{equation}
Therefore GMMG model in non-linear regime has two bulk local degrees of freedom. This is exactly the number of massive graviton which we have obtained in section 3 by a linear analysis. The importance of Hamiltonian analysis is its independence of background. Moreover we see that this model is free of Boulware-Deser ghost.
\section{Conclusions}
In this paper we have generalized recently introduced Minimal Massive Gravity (MMG) model \cite{17} to Generalized Minimal Massive Gravity (GMMG) in asymptotically
$AdS_3$ background. MMG is an extension of TMG, but in contrast to TMG, there is not "bulk vs boundary" clash in the framework of this new model.
Although MMG is qualitatively different from TMG, it has
locally the same structure as that of TMG model. Moreover both models have the same spectrum \cite{18}.
Parallel to this extension, GMMG is an extension of GMG, so one can obtain GMMG  by adding the CS deformation term, the higher derivative deformation term, and an extra term
to pure Einstein gravity with a negative cosmological constant. This last extra term is exactly what the authors of \cite{17} have added to the TMG to obtain
 their interesting model, i.e. MMG. So importance of the work \cite{17} is not only solve the problem of TMG, but also do this by introduction only
  one parameter.\\
 Here we have studied the linearized excitations around the
$AdS_3$ background, and have shown that in contrast to MMG, when GMMG linearized about a $AdS_3$ vacuum, a couple
massive graviton modes appear. At a special, so-called tricritical point in parameter space the two massive graviton solutions become massless and they
are replaced by two solutions with logarithmic and logarithmic-squared boundary behavior.\\
We have calculated the GMMG action to quadratic order about $AdS_3$, and have shown that under condition $\bar{\sigma}+\frac{\gamma }{2\mu^{2}l^2}+\frac{s}{2m^2l^2}>0$, the theory is free of negative-energy bulk modes, so this is the no-ghost condition in the context of GMMG.
We have calculated explicitly the central charges of the CFT dual, using no-ghost condition the right central charge is positive. If $ \bar{\sigma} +\frac{\gamma }{2 \mu ^{2}l^2} + \frac{s }{2m^{2}l^2}>\frac{1}{\mu l} $, then the left central charge is also positive. This condition is not contradict with previous no-ghost condition. So GMMG also avoids the aforementioned ``bulk-boundary unitarity clash''. \\
We have found a critical line, where the left central charge $c_L=0$. In this critical line where massive graviton $m_1$ degenerate
with left-moving massless graviton, a logarithmic mode $h_{mn}^{log}$ appear. Another critical line has obtained when $m_1=m_2$, so
 in this case the operator $D^{m_1}$ and $D^{m_2}$ degenerate. At the intersection of mentioned critical lines, one can see a critical point, where
 three operators $D^{L}$, $D^{m_1}$ and $D^{m_2}$ degenerate, so this critical point is a tricritical point. Similar to this tricritical point
 there is another point in the parameter space of GMMG model where right central charge $c_R=0$, and three operators $D^{R}$, $D^{m_1}$ and $D^{m_2}$ degenerate.
 Due to the presence of these tricritical points, more than logarithmic modes, we obtained the squared-logarithmic modes $h_{mn}^{log^{2}}$ in GMMG.
 These $log$ and $log-$squared modes in contrast with left and right moving massless gravitons do not satisfy the Brown-Henneaux boundary conditions,
 they obey $log$ and $log^2$ asymptotic behavior at the boundary, exactly similar to the corresponding modes in GMG \cite{16}. These arguments support the conjecture
 that GMMG with $log$ and $log^2$ modes, is dual of a rank-3 LCFT.\\
 Here we should mention that the presence of $log$-modes, (or $log^2$ modes) is a sign of non-unitarity, as the theory with modified boundary conditions is expected to be dual to a logarithmic conformal field theory, which are well-known to be non-unitary. Also as we have mentioned in footnote 2, having products of d'Alembertian operators in equation of motion is usually a sign of ghosts, but one thing to keep in mind is that the ghosts arising from the Ostrogradski instability are genuinely different from what is called the Boulware-Deser ghost. This is an additional degree of freedom corresponding to a scalar ghost, which is only manifest in the non-linear theory. So in the  higher order derivative theory, there are massive spin-2 ghosts and Boulware-Deser ghosts. The latter can be removed by tuning the precise coefficients in the action, but the former cannot be removed in a higher-derivative theory of gravity \cite{24,25}.\\
  Then we have investigated the relation between GMMG and GZDG models. We have shown that the GMMG model can be obtained from ZDG plus a Lorentz Chern-Simons term for one of the two spin-connections.\\
  Hamiltonian formulation allows to count the number of local degrees of freedom in the non-linear regime. In section 7 by Hamiltonian analysis we have shown that the Lagrangian 3-form (3) defines a model describing two bulk degrees of freedom. So fortunately GMMG model is free of Boulwar-Deser ghost. But we should mention that situation here is similar to ZDG model. Our model is without Boulwar-Deser ghost only if we demand that the linear combination $(1+\alpha\sigma) e^a - \frac{\alpha}{m^2} f^a$ is invertible. As we have shown in section 6, GZDG with some special parameters can reduce to the GMMG. In the other hand from \cite{15} we know that GZDG in contrast with ZDG is free of Boulwar-Deser ghost at all. The point is this, the  $L_{GZDG}$ of \cite{15} is a combination of $L_{ZDG}(\beta_2=0)$ plus Lorentz-Chern-Simons (LCS)term. But $L_{GZDG}$ in this paper is a combination of $L_{ZDG}(\beta_2\neq0)$ plus LCS term. ZDG model with $\beta_2=0$ is free from Boulwar-Deser ghost, but in the case $\beta_2\neq0$, this model has ghost \cite{27}. If one demand that a linear combination of the dreibeine to be invertible, then ZDG will be free of ghost. This is exactly similar to our model. Therefore GMMG propagate two massive graviton with different masses and is free of  Boulwar-Deser ghost by restriction applied to it.

 \section{Acknowledgments}
I thank  Y. Liu, M. H. Vahidinia, G. Giribet, A. F. Goya, A. J. Routh, H. Adami and especially W. Merbis for helpful discussions and correspondence.

\bigskip



\begin{thebibliography}{9}
\bibitem{1}K. S. Stelle, Phys. Rev. D 16, 953, (1977).
\bibitem{2}S. Deser, R. Jackiw, and G. 't Hooft, "Three-dimensional einstein gravity: Dynamics
of flat space," Ann. Phys. 152, 220 (1984).
\bibitem{3}S. Deser and R. Jackiw, "Three-dimensional cosmological gravity: Dynamics of
constant curvature," Annals Phys. 153, 405 (1984).
\bibitem{4}S. Deser, R. Jackiw and S. Templeton, Annals Phys.
140, 372 (1982) [Erratum-ibid. 185, 406.1988 APNYA,
281,409 (1988 APNYA,281,409-449.2000)].
\bibitem{5}K. A. Moussa, G. Clement and C. Leygnac, Class. Quant. Grav. 20, L277, (2003).
\bibitem{6}E. A. Bergshoeff, O. Hohm and P. K. Townsend, Phys. Rev. Lett. 102, 201301, (2009).
\bibitem{7}J. D. Brown and M. Henneaux, Commun. Math. Phys. 104, 207, (1986).
\bibitem{8}Y. Nutku, Class. Quant. Grav. 10, 2657, (1993); A. Bouchareb and G. Clement,
Class. Quant. Grav. 24, 5581, (2007); A. Maloney, W. Song and A. Strominger, Phys. Rev. D
81, 064007, (2010); D. Anninos, W. Li, M. Padi, W. Song and A. Strominger, JHEP 0903, 130, (2009).
\bibitem{9} A. Sinha, JHEP 1006, 061, (2010) .
\bibitem{10}R. C. Myers and A. Sinha, JHEP 1101, 125, (2011).
\bibitem{11}S. Deser and B. Tekin, Class. Quant. Grav. 20, L259, (2003).
\bibitem{12}I. Gullu, T. C. Sisman and B. Tekin, Class. Quant. Grav. 27, 162001, (2010).
\bibitem{13}M. F. Paulos, Phys. Rev. D 82, 084042, (2010).
\bibitem{14}E. A. Bergshoeff, S. de Haan, O. Hohm, W. Merbis, and P. K. Townsend,
Phys. Rev. Lett. 111, 111102, (2013).
 \bibitem{15}E. A. Bergshoeff, O. Hohm, W. Merbis, A. J. Routh and P. K. Townsend,
Lect. Notes Phys. 892, 181, (2015).
\bibitem{16}Y. Liu, Y. W. Sun, Phys. Rev. D79, 126001, (2009).
\bibitem{17}E. Bergshoeff, O. Hohm, W. Merbis, A. J. Routh and P. K. Townsend,  Class. Quant. Grav. 31, 145008, (2014).
\bibitem{18}A. S. Arvanitakis, A. J. Routh and P. K. Townsend,  Class. Quant. Grav. 31 235012, (2014); A. Baykal, Class. Quant. Grav. 32 (2015) 025013; A. S. Arvanitakis, P. K. Townsend, Class. Quant. Grav. 32(2015) 085003; M. Alishahiha, M. M. Qaemmaqami, A. Naseh, A. Shirzad, arXiv:1409.6146 [hep-th]; G. Giribet, Y. Vásquez, Phys. Rev. D 91, 024026, (2015); M. R. Setare, H. Adami,  	Phys. Rev. D 91, 104039 (2015); A. S. Arvanitakis, Class. Quant. Grav. 32 115010, (2015).
\bibitem{tek1}B. Tekin, arXiv:1503.07488 [hep-th].
\bibitem{19}W. Li, W. Song, A. Strominger, JHEP 0804, 082, (2008).
\bibitem{20}D. Grumiller, O. Hohm,  Phys. Lett. B686, 264, (2010).
\bibitem{21}D. Grumiller and I. Sachs,  JHEP 1003, 012, (2010).
\bibitem{40}Y. Liu, Y. W. Sun,  JHEP 0904, 106, (2009).
\bibitem{set2} M. R. Setare, H. Adami, Phys. Lett. B 744, 280, (2015).
\bibitem{tek2}B. Tekin, Phys. Rev. D 90, 081701 (2014).
\bibitem{22}D. Grumiller, N. Johansson, T. Zojer, JHEP 01, 090, (2011).
\bibitem{23}E. A. Bergshoeff, S. de Haan, W. Merbis, J. Rosseel, T. Zojer, Phys. Rev. D86, 064037, (2012).
\bibitem{26'}E. A. Bergshoeff, A. F. Goya, W. Merbis, J. Rosseel, JHEP 04 (2014) 012.
\bibitem{25'}O. Hohm, A. Routh, P. K. Townsend and B. Zhang, Phys. Rev. D 86, 084035, (2012); A. Routh,  Phys. Rev. D 88, 024022
(2013).
\bibitem{24}T. Nutma, Phys. Rev. D 85, 124040 (2012).
\bibitem{25}H. R. Afshar, E. A. Bergshoeff, W. Merbis, JHEP 08 (2014) 115.
\bibitem{27}M. Ba˜nados, C. Deffayet,  M. Pino, Phys. Rev. D 88, 124016, (2013).
\end{thebibliography}
\end{document}